\documentclass[reprint,9pt,amsmath,amssymb,aps,superscriptaddress,longbibliography
]{revtex4-2}

\usepackage{extarrows}
\usepackage{graphicx}
\usepackage{dcolumn}
\usepackage{bm}
\usepackage[dvipsnames]{xcolor}
\usepackage[]{natbib}
\usepackage{amsmath}

\usepackage{soul}

\mathchardef\mhyphen="2D

\begin{document}
	
	\preprint{APS/123-QED}
	
	\title{Modeling opinion polarization under perception bias}
	
	\author{Hao Yu}
	\affiliation{Research Center for Complexity Sciences, Hangzhou Normal University, Hangzhou,311121, Zhejiang,China}
	
	\author{Bin Xue}
	\affiliation{Research Center for Complexity Sciences, Hangzhou Normal University, Hangzhou,311121, Zhejiang,China}

	\author{Yanpeng Zhu}
	\affiliation{Research Center for Complexity Sciences, Hangzhou Normal University, Hangzhou,311121, Zhejiang,China}
	
	\author{Jianlin Zhang}
	\affiliation{Research Center for Complexity Sciences, Hangzhou Normal University, Hangzhou,311121, Zhejiang,China}
		
	\author{Run-Ran Liu}
	\email{runranliu@163.com}
	\affiliation{Research Center for Complexity Sciences, Hangzhou Normal University, Hangzhou,311121, Zhejiang,China}
	
	\author{Yu Liu}
	\email{yu.ernest.liu@bnu.edu.cn}
	\affiliation{International Academic Center of Complex Systems, Beijing Normal University, Zhuhai, 519087, China}
	
	\author{Fanyuan Meng}
	\email{fanyuan.meng@hznu.edu.cn}
	\affiliation{Research Center for Complexity Sciences, Hangzhou Normal University, Hangzhou,311121, Zhejiang,China}

	%
	

	
	

\begin{abstract}

Social networks have provided a platform for the effective exchange of ideas or opinions but also served as a hotbed of polarization. 
While much research attempts to explore different causes of opinion polarization, the effect of perception bias caused by the network structure itself is largely understudied. To this end, we propose a threshold model that simulates the evolution of opinions by taking into account the perception bias, which is the gap between global information and locally available information from the neighborhood within networks. Our findings suggest that polarization occurs when the collective stubbornness of the population exceeds a critical value which is largely affected by the perception bias. In addition, as the level of stubbornness grows, the occurrence of first-order and second-order phase transitions between consensus and polarization becomes more prevalent, and the types of these phase transitions rely on the initial proportion of active opinions. Notably, for regular network structures, a step-wise pattern emerges that corresponds to various levels of polarization and is strongly associated with the formation of echo chambers. Our research presents a valuable framework for investigating the connection between perception bias and opinion polarization and provides valuable insights for mitigating polarization in the context of biased information.
\end{abstract}


\maketitle

\section{Introduction}
Social networks have emerged as crucial components of modern society, providing individuals with a platform to access and share information, express their opinions, and communicate with each other \cite{jackson2008social,marin2011social,li2013consensus}. Apart from facilitating the aggregation of ideas or opinions \cite{acemoglu2011opinion,buechel2015opinion}, social networks can also reinforce an individual's pre-existing beliefs and attitudes \cite{li2013consensus,pariser2011filter,garrett2009echo,van2021social}, potentially leading to opinions being more polarized \cite{bakshy2015exposure,williams2015network,baldassarri2008partisans}. Opinion polarization is often described as the emergence of divergent views, attitudes, or preferences among individuals or groups towards different directions, leading to opposition and divergence of opinions \cite{sunstein2009going,iyengar2012affect}. The phenomenon of opinion polarization in social networks has become increasingly prevalent in modern society, and it can have significant negative impacts on various aspects of life, including politics, public policy, and even personal relationships. It triggers unfavorable consequences, including conflicts or attacks \cite{dimaggio1996have,mouw2001culture,brady2020mad}, the dissemination of misinformation through social media \cite{iandoli2021impact,zhuravskaya2020political,guerra2017antagonism,sikder2020minimalistic}, and even extremism \cite{nguyen2020bias,luttig2017authoritarianism,stephens2021preventing}. The partisan divisions resulting from opinion polarization not only impede compromise in the design and implementation of social policies but also have far-reaching consequences for the effective function of democracy more broadly \cite{bail2018exposure}. Therefore, it is crucial to investigate the factors that contribute to opinion polarization in social networks.

Numerous studies suggest that homophily, the tendency of individuals to interact primarily with like-minded individuals, is a major contributor to opinion polarization \cite{fenoaltea2023phase,meng2023disagreement,dandekar2013biased,fu2016opinion,bienenstock1990effect}. This effect can lead to the formation of echo chambers \cite{pariser2011filter,petrov2018modeling}, which in turn intensify polarization \cite{barbera2020social,baumann2020modeling}. In addition, confirmation bias\cite{nickerson1998confirmation} or selective exposure, the tendency to accept information that confirms one's existing beliefs and disregard conflicting information \cite{sikder2020minimalistic,dahlgren2020media}, is another significant factor in polarization. Furthermore, the recommendation algorithms of social media platforms like Facebook and Twitter constantly suggest content similar to users' preferences, reinforcing their pre-existing beliefs \cite{van2021social,santos2021link,yarchi2021political,levy2021social}. This narrowing of individual information sources based on specific tastes also contributes to increased polarization. 


Despite the considerable research into the factors that contribute to opinion polarization, there has been limited research on the impact of perception bias \cite{lee2019homophily} on opinion dynamics. Perception bias refers to limited information-gathering abilities based on network structures and can prevent individuals from fully understanding the network's overall information.  This bias can have a significant impact on opinion dynamics, yet there has been no previous attempt to establish and examine the correlation between perceptual bias and opinion polarization in social networks. To address this research gap, we propose a threshold model that simulates the evolution of opinions based on the gap between globally available information and locally accessible information from the neighborhood within networks. Our study reveals that opinion polarization is strongly influenced by the collective stubbornness of the population, particularly when this stubbornness exceeds a critical threshold. This critical threshold is heavily impacted by perception bias. As stubbornness levels increase, first-order and second-order phase transitions between consensus and polarization occur. Most interestingly, for regular networks, we observe a distinct step-wise pattern that corresponds to varying levels of polarization. This pattern is closely linked to the formation of echo chambers.

\section{Model}
A social network consists of $n$ agents, each holding a binary opinion $s_i \in \{0,1\}$ for $i = \{1,2,\dots,n\}$. The degree distribution conforms to $p(k)$. At time step $t=0$, an initial fraction $r$ of the agents adopt an opinion 1 as an active state, while the remaining adopt opinion 0 as an inactive state. Each agent has a \emph{perception bias} $b$, i.e. some level of bias in their perception of the opinions of others, by balancing a global sampling of all agents' opinions with a local sampling of neighborhood opinions. At each subsequent time step $t=1,2,\dots$, an agent with an inactive opinion 0 will adopt an active opinion 1 if their perception of the prevalence of active opinions in the network exceeds some threshold $\phi$. This threshold value can be interpreted as the common \emph{stubbornness} of the population. The dynamic process ends when no agent's opinion changes, meaning that the social network has reached a stable state where all agents are either in an active or inactive opinion.

\begin{figure}[h]
    \centering
    \includegraphics[width=0.9\linewidth]{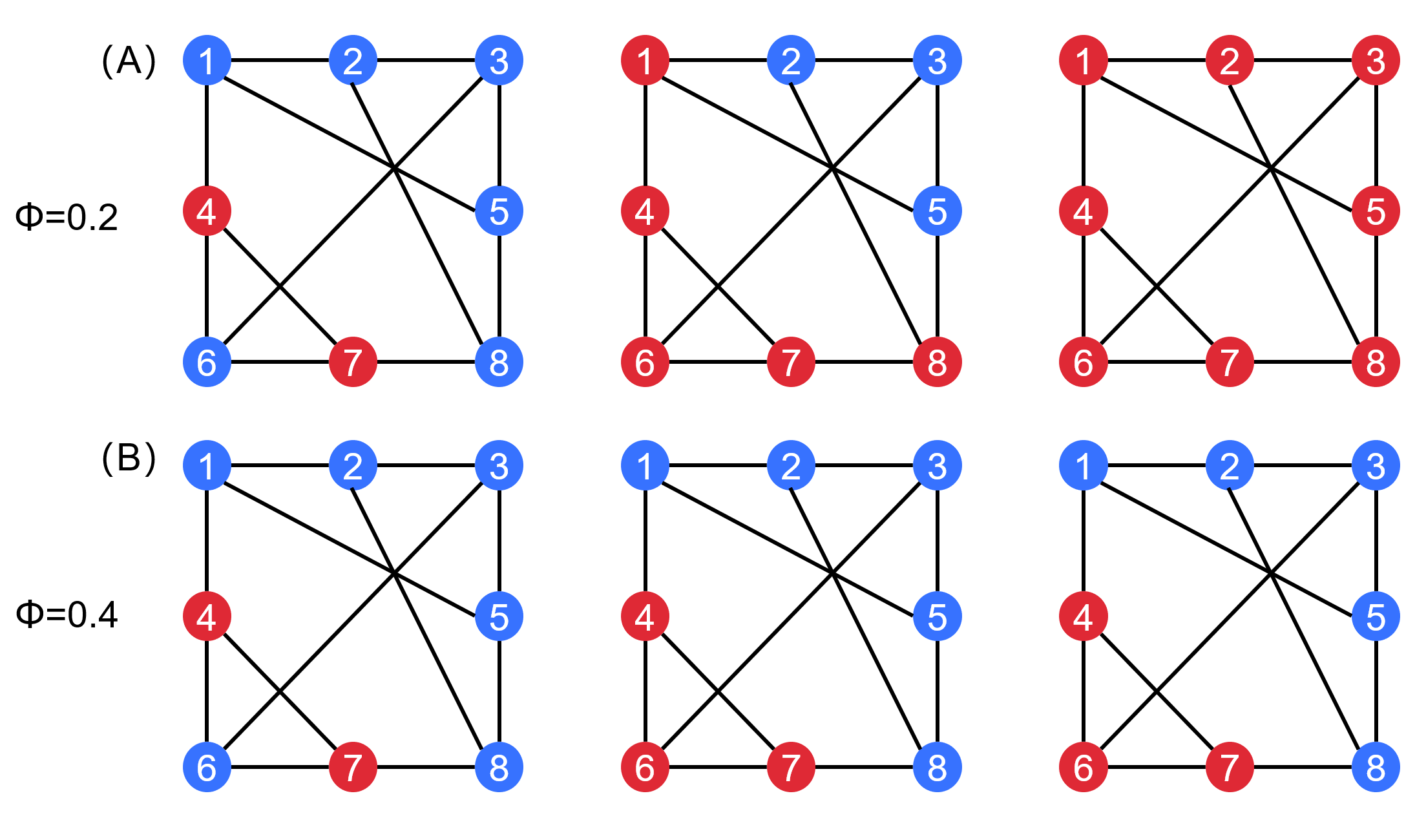}
    \caption{The Figure illustrates the evolution of opinions on a 3-regular network with perception bias $b=1$ for different levels of stubbornness. (A) When the stubbornness parameter $\phi=0.2$ is always lower than the perception of neighborhood active opinions, the population finally reaches a consensus; (B) However, when the stubbornness parameter $\phi=0.4$ becomes greater than the perception of neighborhood active opinions after the second step, some agents with inactive opinions remain unchanged by forming an echo chamber.}
    \label{fig:illu}
\end{figure}

Fig.\ref{fig:illu} illustrates a 3-regular network of $n=8$ agents under perception bias $b=1$, where each agent only considers the opinions of its neighbors. In the beginning, agents 4 and 7 have active opinions, while the rest have inactive opinions. When the stubbornness parameter is set to $\phi=0.2$, the fraction of neighbors with active opinions for agents 1, 6 and 8 are greater than $\phi=0.2$, causing them to update their opinions to active opinions. Then, agents 2 and 5 do, and finally, agent 3 does, resulting in a consensus among all agents (see Fig.\ref{fig:illu}(A)); However, when the stubbornness parameter is set to $\phi=0.4$, the evolution process is similar, but it terminates after the second step because the perception of active opinions for agents 1, 2, 3, 5, 8 never exceeds the stubbornness by forming an echo chamber (see Fig.\ref{fig:illu}(B)).

\section{Results} 
\subsection{The evolution of perception}

For any agent $i$ holding an inactive opinion at time step $t$, the level of perception $p_i(t)$ over active opinions can be expressed by the following form
\begin{equation} \label{eq:pi}
    p_i(t) = (1-b) \frac{\sum_{j=1}^{n} E_{ij} s_j(t)}{n}+ b \frac{\sum_{j=1}^{n} A_{ij} s_j(t)}{k_i},
\end{equation}
where $E$ is an $n \times n$ matrix whose elements are all equal to 1, $A$ is the $n \times n$ adjacency matrix of the network, and $k_i$ is the level of agent $i$. The terms $\sum_{j=1}^{n} E_{ij} s_j(t)/n$ and $\sum_{j=1}^{n} A_{ij} s_j(t)/{k_i}$ can be interpreted as the global fraction of agents and the neighborhood fraction of agent $i$ who hold active opinions, respectively.

Here, we introduce a transition matrix similar to PageRank, which is defined as

\begin{equation} \label{eq:transition_matrix}
T=(1-b)\frac{E}{n} + b D^{-1}A.
\end{equation}

In this equation, $D$ is the degree matrix which contains the degree of each agent along the diagonal entries. From a technical standpoint, the parameter $b$ serves as a controlling factor that regulates the trade-off between the impact of the network structure and a uniform distribution of opinions across all agents. The expression $E/n$ denotes the scenario where each agent assigns equal weights to all opinions within the network, while $D^{-1}A$ represents the scenario where each agent assigns weights to its neighboring opinions proportional to their connectivity.


Thus the perception vector $P(t)=\{p_1(t),p_2(t),\dots,p_n(t)\}$ of all agents over active opinions can be represented as a function of the opinion vector $s(t)=\{s_1(t),s_2(t),\dots,s_n(t)\}$ using the following equation

\begin{equation} \label{eq:motion_perception_bias}
P(t)= Ts(t).
\end{equation}

\subsection{The evolution of polarization}

Since we make the assumption that an agent $i$ with an inactive opinion will adopt an active opinion only when its perception of active opinion exceeds its stubbornness, i.e. $p_i(t) > \phi$, the law of motion for agent $i$'s opinion can be given by

\begin{equation}
s_i(t+1) = s_i(t) + \delta\Big(T(i,:)s(t)\Big)\Big(1-s_i(t)\Big),
\end{equation}

where $\delta(x)$ is the Heaviside step function defined as

\begin{equation}
\delta(x) = \left \{
\begin{aligned}
& 1, \quad x > \phi,\\
& 0, \quad x \leq \phi,
\end{aligned}
\right.
\end{equation}

and $T(i,:)s(t)$ is the dot product of the $i$-th row of matrix $T$ and the opinion vector $s(t)$. Thus the law of motion for all agents' opinions can read as 
\begin{equation}\label{eq:s_motion}
    s(t+1) = 
   \Big(I-\hat{D}(t)\Big)s(t)+ \hat{D}(t) e,
\end{equation}
where \begin{equation}
    \hat{D}(t) = \begin{pmatrix}
  & \delta_1(t) \\
  & &\delta_2(t) \\
  & & & \ddots \\
  & & & & \delta_n(t)
\end{pmatrix}.
\end{equation}
Here $\hat{D}(t)$ is a diagonal matrix with its $i$-th diagonal element $\delta_i(t) = \delta\Big(T(i,:)s(t)\Big)$, and $e$ is a column vector with all elements equal to 1.

At each time step $t$, the fraction of agents with active opinions is $f = \sum_{i=1}^n s_i(t)/n$. Thus the level of polarization $\rho$ can be defined as

\begin{equation}
    \rho = f-0.5.
\end{equation}

The value of $\rho$ lies in the interval $[-0.5,0.5]$. If the absolute value of $\rho$ does not equal 0.5, it indicates that some level of polarization exists among the agents. A higher absolute value of $\rho$ indicates greater polarization, while a lower absolute value of $\rho$ indicates less polarization. When $\rho=0$, it indicates the population reaches a complete polarization.

When the system reaches a stable state at time step $t=\tau$, i.e. $s(\tau+1)=s(\tau)$, by Eq.\ref{eq:s_motion} we can get 

\begin{equation} \label{eq:matrix_equation}
    \hat{D}(\tau) \Big(e-s(\tau)\Big) = 0.
\end{equation}

If there exists some level of polarization (or the population does not reach a consensus) at the stable time step $\tau$,  i.e. $s(\tau) \neq e$, then we must have
\begin{equation} \label{eq:det}
    det(\hat{D}(\tau))=\prod_{i=1}^n \delta_i(\tau)=0.
\end{equation}
In other words, $\delta_i(\tau)=0$ must exist. It indicates that there is still at least one agent $i$ with $s_i(\tau) = 0$ and the following inequality holds
\begin{equation} \label{eq: condition}
        \phi \geq p_i(\tau)= (1-b) \frac{\sum_{j=1}^{n} E_{ij} s_j(\tau)}{n}+ b \frac{\sum_{j=1}^{n} A_{ij} s_j(\tau)}{k_i}.
\end{equation}

To compute the mean level of polarization, we use the computation methods borrowed from the zero-temperature random-field Ising model \cite{gleeson2007seed}. In a stable state, the probability $R$ that a random agent with an active opinion can be reached by another random agent with an inactive opinion, and the probability $\mu_f$ that a random agent stays an active opinion, can be given by the following equations 
\begin{widetext}
			\begin{equation} \label{eq:mu_R}
		\left \{
		\begin{aligned}
			R = r +  (1-r) \sum_{k=1}^\infty \frac{kp(k)}{\langle k \rangle} \sum_{c=0}^{k-1} \binom{k-1}{c} R^c (1-R)^{k-1-c} \delta\Big((1-b) \mu_f+ b\frac{c}{k}\Big), \\
			\mu_f = r + (1-r) \sum_{k=1}^\infty p(k) \sum_{c=0}^{k} \binom{k}{c} R^c (1-R)^{k-c} \delta\Big((1-b) \mu_f+ b\frac{c}{k}\Big).
		\end{aligned} 
		\right.
	\end{equation}
\end{widetext}

The initial condition is $R(t=0)=\mu_f(t=0)=r$.
Thus the mean level of polarization $\mu_{\rho}$ can be computed by $ \mu_{\rho} = \mu_f-0.5$.

\subsection{Opinion polarization without perception bias}

\begin{figure}[ht]
    \centering
    \includegraphics[width=\linewidth]{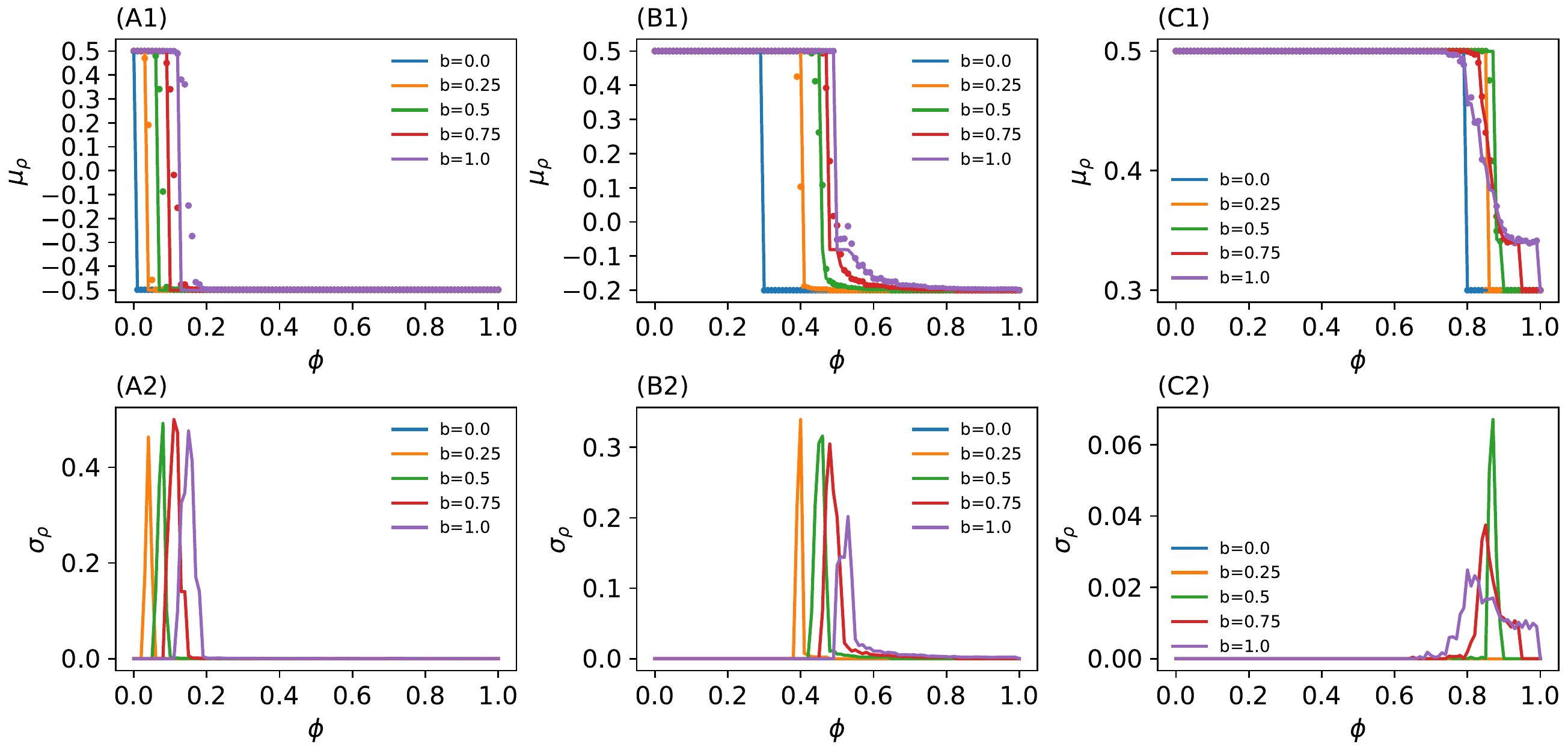}
    \caption{The figure shows the relationship between polarization and stubbornness for various perception biases in Erdős-Rényi random networks, where the markers represent simulation results, and dashed lines represent theoretical predictions. Panels (A1-A2), (B1-B2), and (C1-C2) correspond to $r=0.002$, $r=0.3$, and $r=0.8$, respectively. The simulation settings are as follows: $n=500$, $\langle k \rangle =8$, and each simulation point is an average of 100 model realizations.}
    \label{fig:er1}
\end{figure}

If the perception bias is absent ($b=0$), it means that all agents can aggregate global opinions to make decisions. By using Eqs.\ref{eq:transition_matrix} and \ref{eq:motion_perception_bias}, regardless of time $t$, we can obtain the following expression
\begin{equation}
    P(t) = \frac{Es(t)}{n}=r e.
\end{equation}
There are two possible scenarios to consider inspired by  Eq.\ref{eq:s_motion}. Firstly, if $\phi \leq r$, then $\hat{D}(t) = I$, and we can obtain the value of $s(t+1)=e$. Secondly, if $\phi > r$, then $\hat{D}(t) = 0$, and we can obtain the value of $s(t+1)=0$. They imply that the level of stubbornness among the agents is the only obstacle to achieving consensus, regardless of the network structure. Hence, we can express the critical level of stubbornness $\phi^I_c$ as follows
\begin{equation}
    \phi^I_c = r.
\end{equation}
If the stubbornness exceeds the critical value $\phi^I_c$, it means that agents with inactive opinions are too stubborn to switch to active opinions, then the level of polarization depends solely on the initial fraction $r$ of agents with active opinions, specifically, $\rho=r-0.5$. Therefore, the greater the difference between $r$ and 0.5, the more polarized the population will be. By contrast, the stubbornness is lower than the critical value $\phi^I_c$, individuals are open to changing their opinions, and the population will eventually converge to a consensus (see Figure \ref{fig:er1}).

\subsection{Opinion polarization under perception bias}
\subsubsection{Benchmark: Erdős-Rényi networks}
Let us first consider the level of polarization in Erdős-Rényi random networks with Poisson degree distribution, i.e. $p(k)=e^{-\langle k \rangle}{\langle k \rangle}^{k}/{k!}$. By Eq.\ref{eq:mu_R}, we can obtain these two probabilities $R$ and $\mu_f$ as
\begin{widetext}
	\begin{equation} \label{eq:mu_R_er}
		\left \{
		\begin{aligned}
			R = r +  (1-r) \sum_{k=1}^\infty \frac{e^{-\langle k \rangle}{\langle k \rangle}^{k-1}}{(k-1)!}  \sum_{c=0}^{k-1} \binom{k-1}{c} R^c (1-R)^{k-1-c} \delta\Big((1-b) \mu_f+ b\frac{c}{k}\Big), \\
			\mu_f = r + (1-r) \sum_{k=1}^\infty \frac{e^{-\langle k \rangle}{\langle k \rangle}^{k}}{k!}\sum_{c=0}^{k} \binom{k}{c} R^c (1-R)^{k-c} \delta\Big((1-b) \mu_f+ b\frac{c}{k}\Big),
		\end{aligned} 
		\right.
	\end{equation}
\end{widetext}

with initial condition $R(t=0)=\mu_f(t=0)=r$.

Using Equation \ref{eq:mu_R_er}, we define an auxiliary equation as 
\begin{widetext}
	\begin{equation} \label{eq:h_R}
		h(R) = r +  (1-r) \sum_k \frac{e^{-\langle k \rangle}{\langle k \rangle}^{k-1}}{(k-1)!}   \sum_{n}^{k-1} \binom{k-1}{n} R^n (1-R)^{k-1-n} \delta\Big((1-b) \mu_f+ b\frac{n}{k}\Big) - R
	\end{equation}
\end{widetext}

to explore the critical values of parameters as phase transitions occur in our model. 
\begin{figure}[ht]
    \centering
    \includegraphics[width=\linewidth]{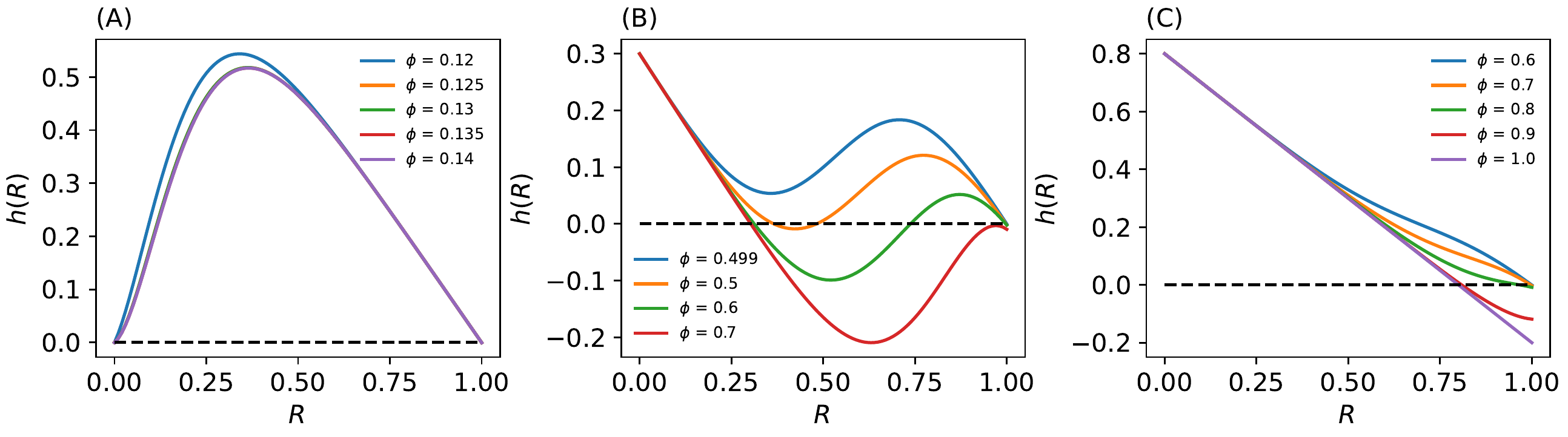}
    \caption{The relation between $h(R)$ with $R$ for different levels of stubbornness in Erdős-Rényi random networks, where $\langle k \rangle =8$, and $b=1$. Panels (A), (B), (C) correspond to different fractions of initial active opinions, i.e. 
    $r=0.002, r=0.3, r=0.8$, respectively.}
    \label{fig:er_R_h_R}
\end{figure}

When the perception bias is at its maximum ($b=1$), agents with inactive opinions rely only on the opinions of their neighbors to decide whether to update their own opinions or not. In this case, our model reduces to a generalization of the Watts Threshold Model \cite{watts2002simple} with an arbitrary fraction of initial active opinions. 

When $r \ll 1$, a first-order phase transition occurs, and the critical level of stubbornness $\phi^I_c$ can be approximated as 
\begin{equation}
    \phi^I_c \approx 1/\langle k \rangle,
\end{equation}
which can be seen in Fig.\ref{fig:er1}(A1-A2) and be comfirmed by Fig.\ref{fig:er_R_h_R}(A).

However, when $r$ is not negligible, the situation becomes more complex, and we must consider two scenarios: $r<0.5$ and $r\geq 0.5$. 

One example is $r=0.3$, as the stubbornness $\phi$ decreases, the level of polarization gradually increases. Then, a first-order phase transition occurs at $\phi^I_c \approx 0.5$, where the polarization level jumps from a higher level to full consensus (shown in Fig.\ref{fig:er1}(B1-B2) and confirmed by Fig.\ref{fig:er_R_h_R}(B)), using the first crossing point between $y=0$ and $h(R)$ in Eq.\ref{eq:h_R} (i.e., $R_c \approx 0.3653$), which yields $\mu_{\rho} \approx -0.08$. Furthermore, as the perception bias decreases, $\phi^I_c$ also decreases. A higher perception bias leads to a higher level of polarization, indicating that polarization is more likely to occur when perception bias is higher.

Another example is $r=0.8$. As the stubbornness $\phi$ decreases, there is a first-order phase transition at $\phi^I_c=1$, where the polarization jumps from 0.3 to around 0.34. Following this, polarization increases and then undergoes a second-order phase transition at $\phi_c \approx 0.8$, jumping to approximately 0.5 (shown in Fig.\ref{fig:er1}(C1-C2) and confirmed by Fig.\ref{fig:er_R_h_R}(C)). Additionally, as the level of perception bias decreases, the critical value $\phi^I_c$ also decreases while $\phi^{II}_c$ increases. This suggests that a critical value $b_c$ exists, below which a higher perception bias leads to a higher level of polarization, while above it, a higher perception bias results in a lower level of polarization.

\begin{figure}[ht]
    \centering
    \includegraphics[width=\linewidth]{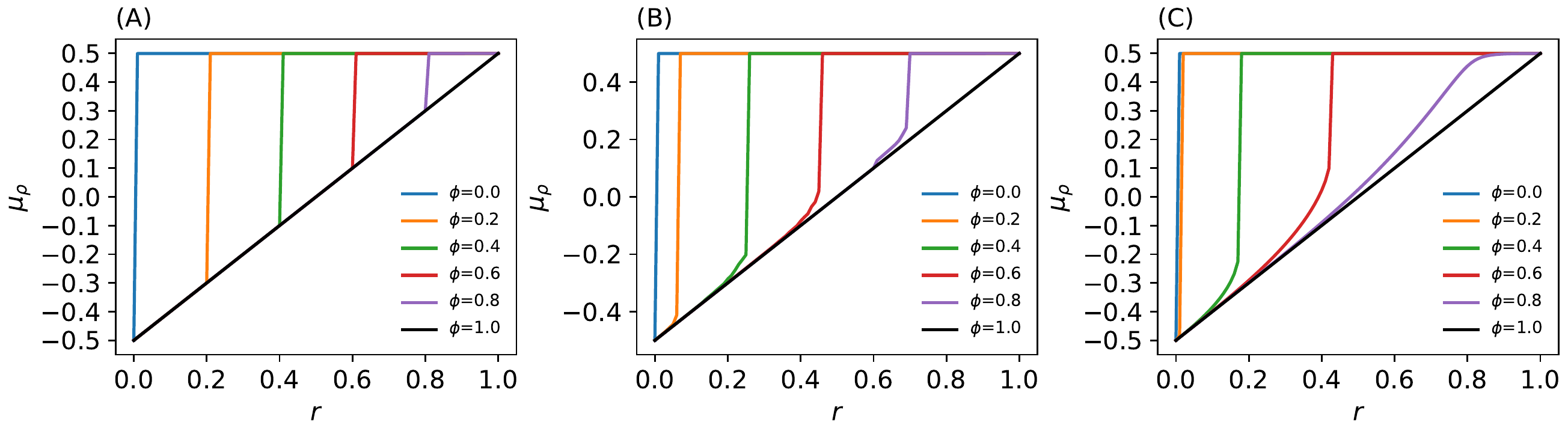}
    \caption{The figure shows the relationship between polarization and stubbornness for different levels of stubbornness in Erdős-Rényi random networks, where $k=8$. Panels (A), (B), and (C) correspond to different levels of perception bias, namely $b=0$, $b=0.5$, $b=1$, respectively.}
    \label{fig:er2}
\end{figure}
\begin{figure}[ht]
    \centering
    \includegraphics[width=\linewidth]{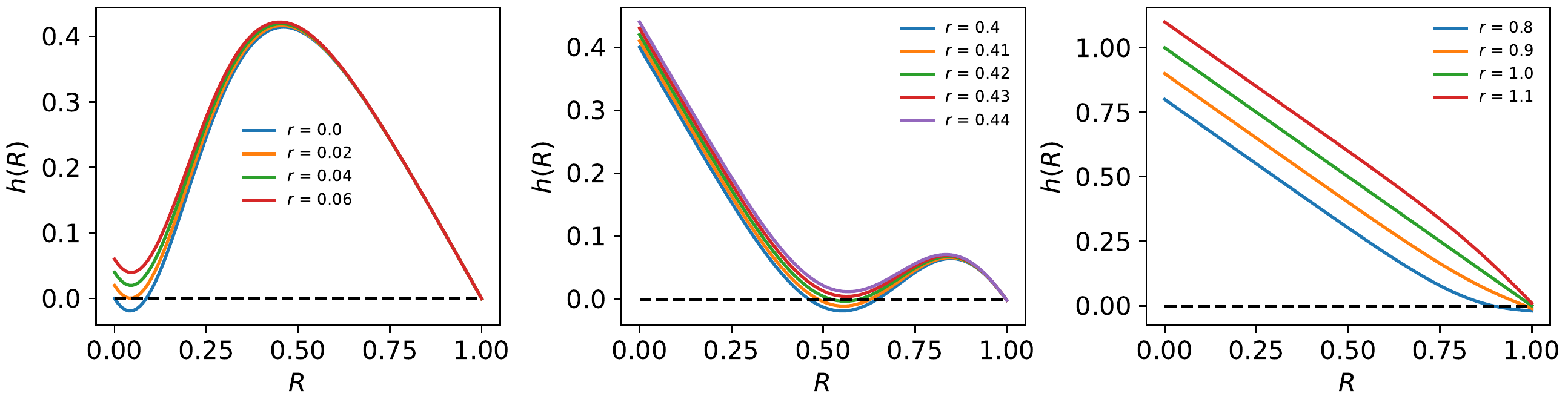}
    \caption{The figure shows the relationship between $h(R)$ and $R$ for different values of $r$ in Erdős-Rényi random networks, where $k=8$ and $b=1$. Panels (A), (B), and (C) correspond to different levels of perception bias, i.e. $b=0$, $b=0.5$, $b=1$, respectively.}
    \label{fig:er_h_R_v2}
\end{figure}

Next, we discuss how $r$ can compensate for polarization when stubbornness is held constant. When the level of stubbornness is at its maximum value ($\phi=1$), the polarization level is determined solely by the parameter $r$, where $\rho=r-0.5$. However, when $\phi \neq 1$, a first-order phase transition may occur as $r$ decreases, regardless of the level of perception bias. For example, in Fig. \ref{fig:er2} and confirmed by Fig. \ref{fig:er_h_R_v2}, the transition occurs at approximately $r^I_c \approx 0.42$ for $\phi=0.6$ and $b=1$. This indicates that when $r>r_c$, the values of $r$ and $\phi$ can compensate for each other to some degree, maintaining the population in a consensus state. Conversely, when $r\leq r_c$, the polarization level may increase and then decrease until it is determined solely by $r$.


\subsubsection{Extension: $k$-regular networks}
For $k$-regular networks, we have $p(k)=1$. Then Eq.\ref{eq:mu_R} can be simplified to 
\begin{widetext}
\begin{equation}\label{eq:mu_R_rg}
\left \{
\begin{aligned}
        R = r +  (1-r) \sum_{c=0}^{k-1} \binom{k-1}{c} R^c (1-R)^{k-1-c} \delta\Big((1-b) \mu_f+ b\frac{c}{k}\Big), \\
    \mu_f = r + (1-r) \sum_{c=0}^{k} \binom{k}{c} R^c (1-R)^{k-c} \delta\Big((1-b) \mu_f+ b\frac{c}{k}\Big),
\end{aligned} 
\right.
\end{equation}
\end{widetext}
with initial condition $R(t=0)=\mu_f(t=0)=r$. 

\begin{figure}[ht]
    \centering
\includegraphics[width=\linewidth]{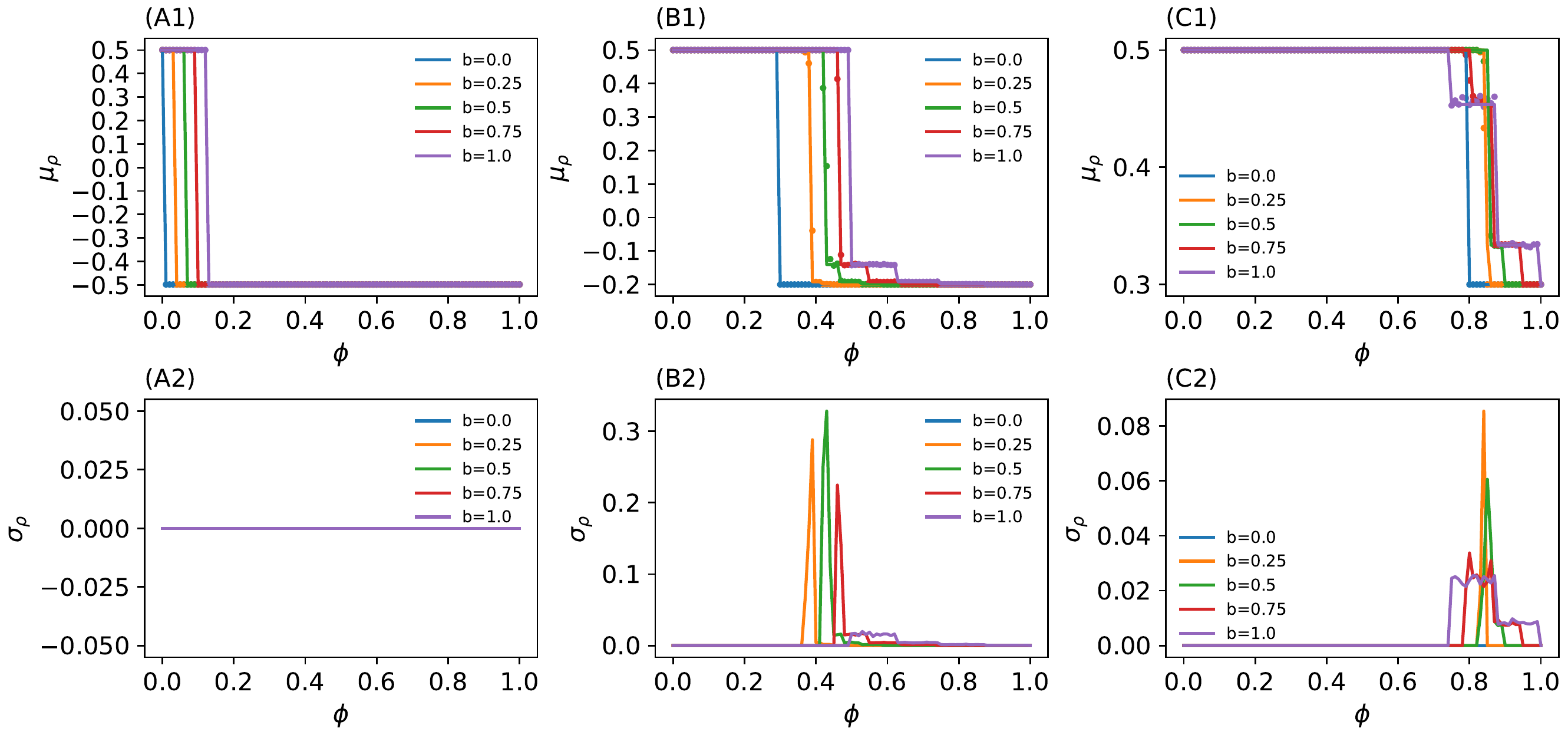}
    \caption{The figure shows the relationship between polarization and stubbornness for different perception biases in $k$-regular networks, where markers represent simulation results and dashed lines represent theoretical predictions. anels (A1-A2), (B1-B2), and (C1-C2) correspond to $r=0.002$, $r=0.3$, and $r=0.8$, respectively. The simulation settings are as follows: $n=500$, $\langle k \rangle =8$, and each simulation point is an average of 100 model realizations.}
     \label{fig:rg}
\end{figure}

We can observe that the mean level of polarization $\mu_{\rho}$ as a function of the threshold $\phi$ always exhibits a step-wise pattern, regardless of the level of perception bias (see Fig.\ref{fig:rg}). 

If the perception bias is maximum ($b=1$), from Eq.\ref{eq:mu_R_rg}, we can easily get the critical level of stubbornness as 
\begin{equation}
    \phi^I_c = \frac{i}{k}, \{i=1,2,\dots,k\}
\end{equation}
which can be seen in Fig.\ref{fig:rg}. 


If the perception bias lies in the range $b\in (0,1)$, the lower and upper bounds of $\phi^I_c$ for the step-wise stable state of polarization $\mu_f^*$ can be given by
\begin{equation} \label{eq:phi_range_general}
    b/k+(1-b)\mu_f^*\leq \phi^I_c \leq b+(1-b) \mu_f^*.
\end{equation}

\newpage
\section{Conclusion and Discussion}

In order to effectively combat the negative effects of opinion polarization in social networks, it is crucial to have a deep understanding of the various factors that contribute to it. Previous research has identified several key factors, such as homophily, confirmation bias, and recommendation algorithms. However, there is one important factor that has often been overlooked in these studies: perception bias.

Perception bias can have a significant impact on the dynamics of opinion formation and polarization, yet it has received relatively little attention in previous research. Our proposed threshold model sheds light on the relationship between perception bias and opinion polarization, underscoring the importance of addressing this factor in efforts to mitigate polarization.

Despite its valuable contributions, our model does have some limitations. For instance, it simplifies the structure of real-world networks by utilizing generated networks, and it assumes that each agent possesses unique levels of perception and personal stubbornness. Additionally, it does not take into account external factors that can influence opinion formation and polarization. Moving forward, further research can explore these aspects in order to develop more comprehensive models.

Overall, our study underscores the importance of addressing both perception bias and collective stubbornness in efforts to combat opinion polarization and highlights the need for more accurate and nuanced models that can capture the complexities of real-world scenarios. By deepening our understanding of the factors that contribute to opinion polarization, we can take meaningful steps toward building a more informed and cohesive society.

\newpage

\bibliographystyle{unsrt}
\bibliography{biase.bib}
\end{document}